 \documentclass[12pt]{article}
\usepackage{graphicx}
\usepackage{pict2e}
\usepackage{dcolumn}
\usepackage{bm}
\usepackage{comment}
\usepackage{rotating}
\usepackage{longtable}
\usepackage{float}
\usepackage{eucal}
\usepackage{csquotes}
\usepackage{xcolor} 

\usepackage{authblk}
\usepackage{longtable}

\makeatletter

\newcommand{\Rmnum}[1]{\expandafter\@slowromancap\romannumeral #1@}
\makeatother

\begin{document}




\title{Coulomb breakup  of neutron-rich $^{29,30}$Na isotopes near the island of inversion}

\author[1]{A . Rahaman}
\author[1,2]{ Ushasi Datta\thanks{corresponding:ushasi.dattapramanik@saha.ac.in}}
\author[2,3]{T. Aumann}
\author[4]{S. Beceiro-Novo}
\author[2]{K. Boretzky}
\author[2]{C. Caesar}
\author[5]{B.V. Carlson}
\author[6]{W.N. Catford}
\author[1]{S. Chakraborty}
\author[7]{M. Chartier}
\author[4]{D. Cortina-Gil}
\author[8]{G. De. Angelis}
\author[2,9]{D. Gonzalez-Diaz}
\author[2]{H. Emling}
\author[4]{P. Diaz Fernandez}
\author[10]{L.M. Fraile}
\author[2]{O. Ershova}
\author[2,11]{H. Geissel}
\author[12]{B. Jonson}
\author[12]{H. Johansson}
\author[13]{N. Kalantar-Nayestanaki}
\author[14]{R. Kr\"ucken}
\author[14]{T. Kr\"oll}
\author[2]{J. Kurcewicz}
\author[2]{C. Langer}
\author[14]{T.Le Bleis}
\author[2]{Y. Leifels}
\author[2]{G. M\"unzenberg}
\author[2]{J. Marganiec}
\author[12]{T. Nilsson}
\author[2]{C. Nociforo}
\author[13]{A. Najafi}
\author[2]{V. Panin}
\author[3]{S. Paschalis}
\author[2]{R. Plag}
\author[2]{R. Reifarth}
\author[13]{C. Rigollet}
\author[2]{V. Ricciardi}
\author[2]{D. Rossi}
\author[3]{H. Scheit}
\author[2]{H. Simon}
\author[2,11]{C. Scheidenberger}
\author[2]{S. Typel}
\author[7]{J. Taylor}
\author[2]{Y. Togano}
\author[3]{V. Volkov}
\author[2]{H. Weick}
\author[15]{A. Wagner}
\author[2]{F. Wamers}
\author[2]{M. Weigand}
\author[2]{J.S. Winfield}
\author[15]{D. Yakorev}
\author[2]{M. Zoric}

\affil[1]{Saha Institute of Nuclear Physics, Kolkata 700064, India}
\affil[2]{GSI Helmholtzzentrum fur Schwerionenforschung GmbH, D-64291 Darmstadt, Germany}
\affil[3]{Technische Universitat Darmstadt, 64289 Darmstadt, Germany}
\affil[4]{Universidade de Santiago de Compostela, 15706 Santiago de Compostela, Spain}
\affil[5]{Instituto Tecnologico de Aeronautica, Sao jose dos Campus, Brazil}
\affil[6]{University of Surrey, Guildford GU2 5XH, United Kingdom}
\affil[7]{University of Liverpool, Liverpool L69 7ZE, United Kingdom}
\affil[8]{INFN, Legnaro, Italy}
\affil[9]{Zaragoza University, 50009 Zaragoza, Spain}
\affil[10]{Universidad Complutense de  Madrid, CEI Moncloa, E-28040 Madrid, Spain}
\affil[11]{II. Physikalisches Institut, D-35392 Giessen}
\affil[12]{Fundamental Fysik, Chalmers Tekniska Hogskola, S-412 96 Goteborg, Sweden}
\affil[13]{KVI-CART, University of Groningen, Groningen, The Netherlands}
\affil[14]{Physik Department E12, Technische Universit\"at M\"unchen, 85748 Garching, Germany}
\affil[15]{Helmholtz-Zentrum Dresden-Rossendorf, D-01328 Dresden, Germany}

\date{\today}

\maketitle

\begin{abstract}
First results are reported  on the ground state configurations of the neutron-rich $^{29,30}$Na isotopes, obtained via  
Coulomb dissociation (CD) measurements as a method of the direct probe. The  invariant mass spectra of those nuclei have been 
obtained through  measurement of  the four-momenta of all decay products after Coulomb excitation on a  $^{208}$Pb target at
energies  of 400-430 MeV/nucleon using FRS-ALADIN-LAND setup at GSI, Darmstadt. Integrated Coulomb-dissociation cross-sections  (CD) 
 of  89 $(7)$ mb and 167 $(13)$ mb up to excitation energy of 10 MeV  for one neutron removal from $^{29}$Na and $^{30}$Na respectively,
 have been extracted. The major part of one neutron removal,  CD cross-sections  of those nuclei populate core, in its' ground state.
A comparison  with the direct breakup model,  suggests the predominant occupation of the valence neutron in the ground state of 
 $^{29}$Na${(3/2^+)}$ and $^{30}$Na${(2^+)}$ is  the $d$ orbital with small contribution in the $s$-orbital  which are   coupled with 
ground state of the core. The  ground state configurations of these nuclei are  as $^{28}$Na$_{gs}(1^+)\otimes\nu_{s,d}$ and 
$^{29}$Na$_{gs}(3/2^+)\otimes\nu_{ s,d}$, respectively. The ground state spin and  parity of these nuclei, obtained from this experiment
 are in agreement with earlier reported values. The spectroscopic factors for the valence neutron occupying the $s$ and $d$ 
orbitals for these nuclei in the ground state have been  extracted and reported for the first time. A comparison of the experimental 
findings with the shell model calculation using MCSM suggests a lower limit of around 4.3 MeV of the sd-pf shell gap in $^{30}$Na. 
\end{abstract}

\vspace{2pc}

\section{Introduction}
\label{}

The magic numbers  \cite{my49,jn49} of the nuclei are a benchmark of nuclear structure.
The underlying shell gap is the characteristic of the mean nuclear field  which
consists of many ingredients of the nucleon-nucleon interactions. 
The modification in the shell gaps through the effects such as the tensor component of the NN force become 
pronounced with large neutron-proton asymmetries in the exotic nuclei far away from stability.
 These lead to the disappearance of established magic numbers and the appearance of new ones.
The first observation of the disappearance of the magic number (N=20) was reported,  based on the mass measurements in  
the neutron rich $^{31,32}$Na  \cite{Ref a}. The experimental observation of the higher binding energies 
 of these nuclei is a direct consequence of the large deformation \cite{Ref a}.  Later large deformation was also reported in
the ground state of $^{32}$Mg \cite{Ref b}. This large deformation was explained by considering the intruder effects 
which suggests a clear vanishing of the shell gap  between $sd$ and $pf$ shell around  $N$ = $20$. 
The $N$ = $20$ isotones with Z $\sim 10-12$ are considered to belong to the ``island of inversion''  
\cite{Ref warburton} where the intruder configurations  dominate the ground state wave function.
Recently, it has been observed that the shell gaps at $N$ = $20$ and $28$  are significantly reduced in the ground 
state of $^{33}$Mg \cite{datt16}.
Otsuka et al. considered  strongly attractive monopole interaction of the tensor  force  to describe the shell evolution for 
several nuclei \cite{Ref otsuka}. The attractive $T$ = $0$ monopole interaction between the 
$\pi$$d_{5/2}$ and $\nu d_{3/2}$ orbits changes the size of the $N$ = $20$ effective energy gap as the protons fill the $d_{5/2}$ orbitals.\\
The availability of radioactive ion beam provides an unique opportunity to study the evolution  of the shell structure of the nuclei
far away from the $\beta$-stability line and many experimental observations on the non magicity behavior of 
the neutron-rich nuclei around so-called magic number have been reported.   It is of particular interest to
 understand the shell evolution for the nuclei where transition from the normal ground state configuration
to the intruder dominated  ground state configuration occurs. The experimental studies in this direction may provide a stringent test for
the validation of various theoretical predictions of the nucleon-nucleon interactions. In other words, to understand change of
 nucleon-nucleon interaction, with iso-spin quantum number, the key ingredients are experimental information on ground state configuration 
of these transition nuclei.  So far several studies have been performed  using different techniques to investigate this region 
\cite{Ref utsuno,Ref c,Ref hurst,mach,Ref tripathi1,Ref tripathi2,doo13}. Though it is established that  valence neutron(s) 
in the ground state of the neutron-rich Na, Mg, Ne isotopes at $N$ = $20$, are occupying $pf$ intruder orbitals, but it is not well
 established for the neighboring nuclei. The ground state configuration of transitional nuclei from normal 
to ``island of inversion'' and experimental information for  the transitional nuclei  are often contradictory to each other. 
Terry et al. \cite{terry} observed $pf$ orbital occupation of valence neutron in the ground state of  neutron-rich $^{28,30}$Ne 
($N$ = $18$, $20$) via  knockout measurements. However, the situation is different  for $^{30}$Mg (N=18). Both knockout \cite{mg1} and
 Coulomb excitation data \cite{mg2} are in agreement  with $sd$-shell ground state configuration of this nucleus.
 No detailed ground state configurations of $^{29,30}$Na ($N$ = $18$, $19$)  is available in the literature.    
 The low-energy level structure of the exotic   $^{28,29}$Na isotopes have been investigated through 
$\beta$-delayed $\gamma$ spectroscopy \cite{Ref tripathi1} and the authors proposed that around $\sim$ 46 $\%$ intruder configuration
  in the ground state of $^{29}$Na \cite{Ref utsuno} is necessary to explain experimental data. 
Several theoretical model calculations and experimental results \cite{Ref utsuno, Ref c, Ref hurst, Ref tripathi1,Ref tripathi2,isol} suggest 
a significant reduction  of the $sd-pf$ shell gap and the ground state of  $^{29,30}$Na are   dominated by intruder states.
 $^{30}$Na has been investigated using Coulomb excitation at the intermediate energies
\cite{Ref c} and  transition probability $B(E2: 2^+\rightarrow 3^+) = 147(21)$ $e^2$ $fm^4$ was reported.
 The Knockout reaction by Tajes et al. \cite{taj} showed that the momentum distributions of both $^{29,30}$Na are almost identical  
(137 MeV/c and 130 MeV/c respectively) although $^{29}$Na (S$_n$ = 4.4 MeV) is much deeply bound than $^{30}$Na (S$_n$ = $2.27$ MeV). 
This unconventional experimental observation  encourage for more experimental investigation.
 An experimental program (GSI:s306) has been initiated to explore the ground state configurations of the  neutron-rich nuclei
 around N $\sim$ 20 via direct probes at GSI, Darmstadt. The Coulomb break up is a direct method to probe the quantum number 
of the valence nucleon of loosely  bound nuclei \cite{ref e, Ref f,udp,dat2}. 
In this article, first results on wave-function decompositions of  the ground state of $^{29,30}$Na, studied  via Coulomb breakup,   
are being reported.

\section{Experiment}
\label{}

\begin{figure}[h]
\centering
\includegraphics[width=9cm]{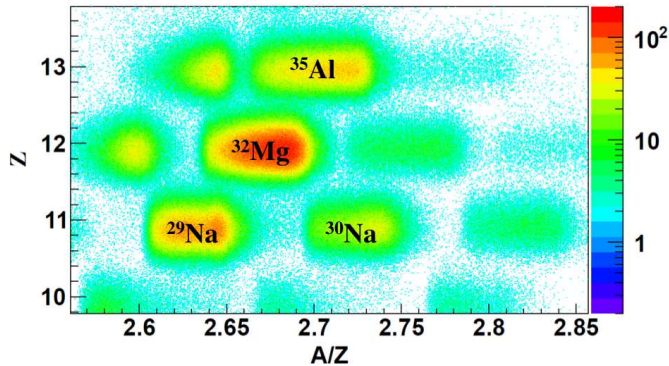}
\caption{Identification plot for mixed radioactive beam impinging on secondary targets.}
\label{inc1}
\end{figure}

 The secondary beam containing  $^{29,30}$Na isotopes along with others were populated by fragmentation of the $^{40}$Ar
 primary beam with energy 540 MeV/nucleon and separated  at FRS \cite{frs}.  The incoming projectiles were 
 identified event-by-event  by measuring the magnetic rigidity,  time of flight and relative energy loss of the exotic nuclei.
The incoming beam-identification plot  is shown in Figure \ref{inc1}. The beam intensity of $^{29,30}$Na were around 14$\% , 
5.5\%$ respectively, of total incoming beam.

\begin{figure}[h]
\centering
\includegraphics[width=9cm]{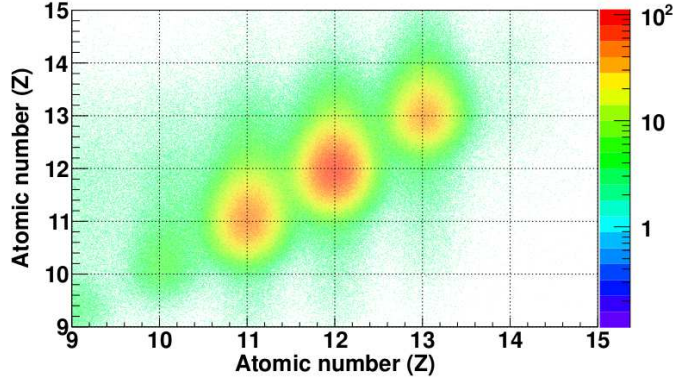}
\caption{Outgoing fragments charge distribution}
\label{outz}
\end{figure}

\begin{figure}[h]
\centering
\includegraphics[width=9cm]{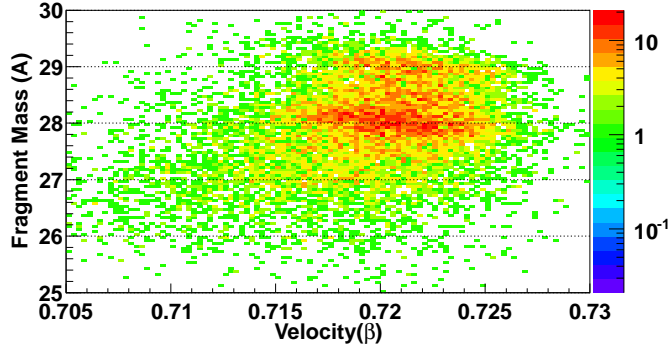}

\caption{Outgoing fragments mass identification plot after one neutron breakup of $^{29}$Na  beam on a lead target.}
\label{outmass}
\end{figure}

 The secondary beam  (Figure \ref{inc1}) was  transported to the experimental site, 
a neighboring cave C where the complete kinematic measurements were performed using  the FRS-ALADIN-LAND setup.
 The position of the  incoming beam before the secondary target and  the position of the reaction fragments after the target 
 were accurately measured using double sided Silicon strip detectors (DSSD). The target was surrounded by
162 NaI(Tl) detectors \cite{metag}, which cover almost $4\pi$ solid angle. This detector-array was used for detecting the
 $\gamma$-rays from the excited core of the projectile after the Coulomb breakup.
 After reaction at the secondary target the reaction fragments as well as the unreacted beam were bent by a Large Dipole 
Magnet (ALADIN) and ultimately were detected at the time of flight
wall detector (TFW) via two scintillator detectors  GFI \cite{gfi1,gfi}. Outgoing fragment charge distribution
can be obtained  through the energy loss at TOF and silicon strip detector (DSSD). Figure \ref{outz} shows the charge distribution
of out going fragments, $Z_{TFW}$, obtained from energy loss at TFW  against $Z_{DSSD}$, obtained from measurement of energy loss at DSSD.
 The decay product, neutrons were forward focused due to Lorentz boost and  detected by the Large Area 
Neutron Detector (LAND) \cite{land}  for the time of flight and position measurements.
 The reaction fragments were bent at different angles inside ALADIN depending on their  charge to momentum ratios.
 This relative deflection angle was measured using the GFI detectors placed at two different distances at an angle of $15^0$ from the
original beam direction after the ALADIN. The mass of  the outgoing reaction fragments was reconstructed
using the deflection angles measured from GFI, the energy loss at TFW, and the time of flight measurement
of the reaction fragments. Figure \ref{outmass} represents the  mass  of the outgoing fragment against speed, after one neutron breakup
of $^{29}$Na.   For details of the experimental setup and detector calibration
 see \cite{anisur, ceser, datt16} and references therein.

\section{Analysis}
\label{}

In this experiment the excitation energy $E^*$ of  $^{29,30}$Na  are  determined by measuring four momenta of all the decay products
of those nuclei  after breakup \cite{ref e, Ref f}.  


\begin{equation}
E^*=\sqrt{(m_f^2+m_n^2)c^4 + 2 \gamma_f \gamma_n m_f m_nc^4(1-\beta_f \beta_n cos(\theta_{fn})) }
\\+E_\gamma - m_{proj}c^2
\end{equation}

In the above equation $m_f$, $m_n$, $m_{proj}$ are the  masses of the breakup products i.e. reaction fragment, neutron  and the projectile respectively.
$\beta_f$, $\beta_n$, $\gamma_f$, $\gamma_n$ are the speeds and  Lorentz factors of the reaction fragment, neutron   respectively.
 $\theta_{fnj}$ represents the angle between the reaction products i.e. reaction fragment and the breakup neutron in the present experiment.
$E_{\gamma}$ is the excitation of the core/fragment of the projectile measured with the help of the crystal ball detector.

\begin{figure}[h]
\centering
\includegraphics[width=9cm]{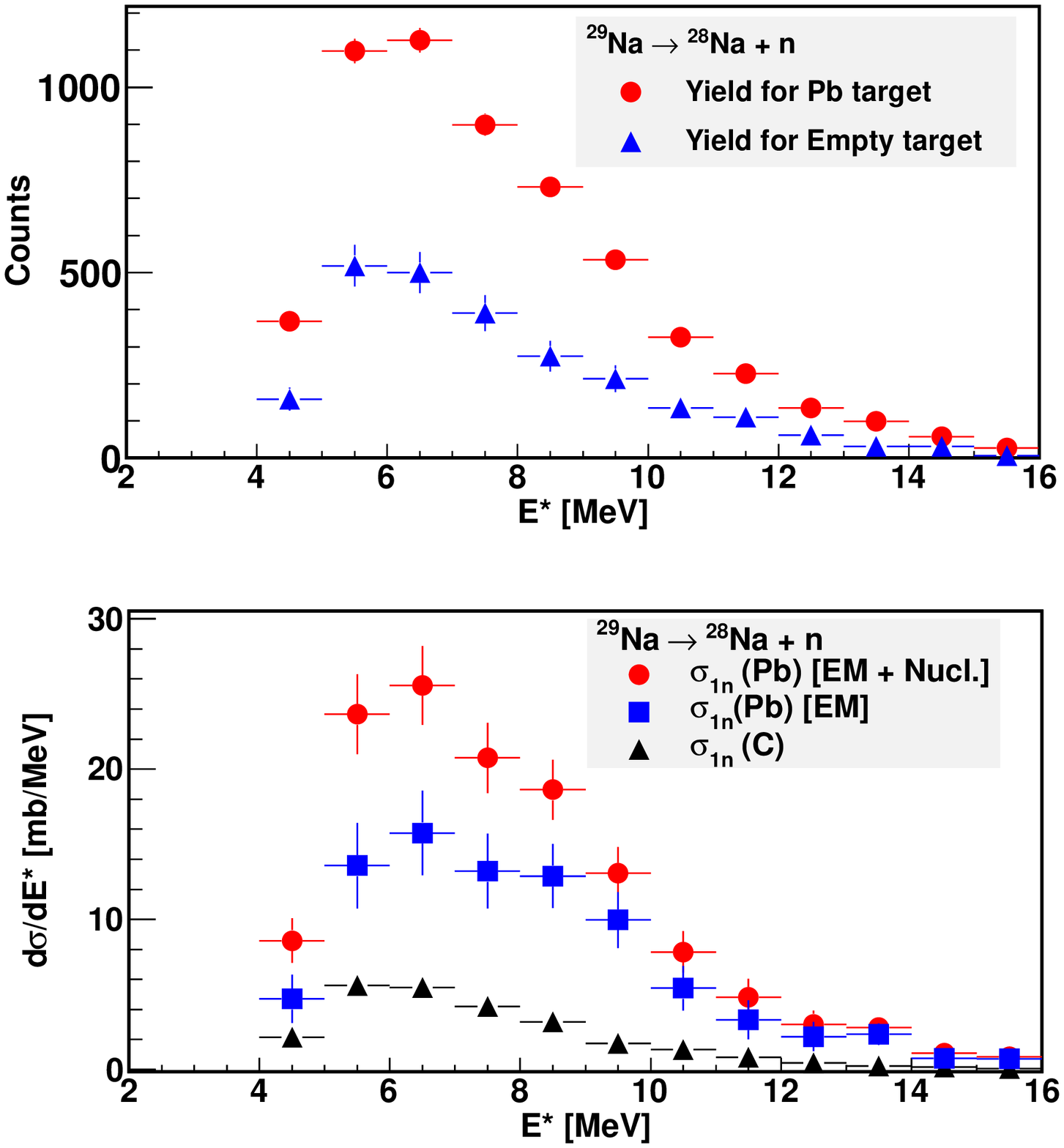}
\caption{(Top) The yields of one neutron breakup reaction against excitation energy  of $^{29}$Na  using Pb target and without any target   
have been represented by filled circles and  triangles, respectively.
(bottom) One neutron breakup cross-sections against excitation energy of $^{29}$Na  using  Pb-target (Coulomb+nuclear), C-target(nuclear) and
 pure Coulomb of Pb have been denoted by circles, triangles and squares, respectively.}
\label{29na} 
\end{figure}

\begin{figure} 
\centering
 \includegraphics[width=9cm]{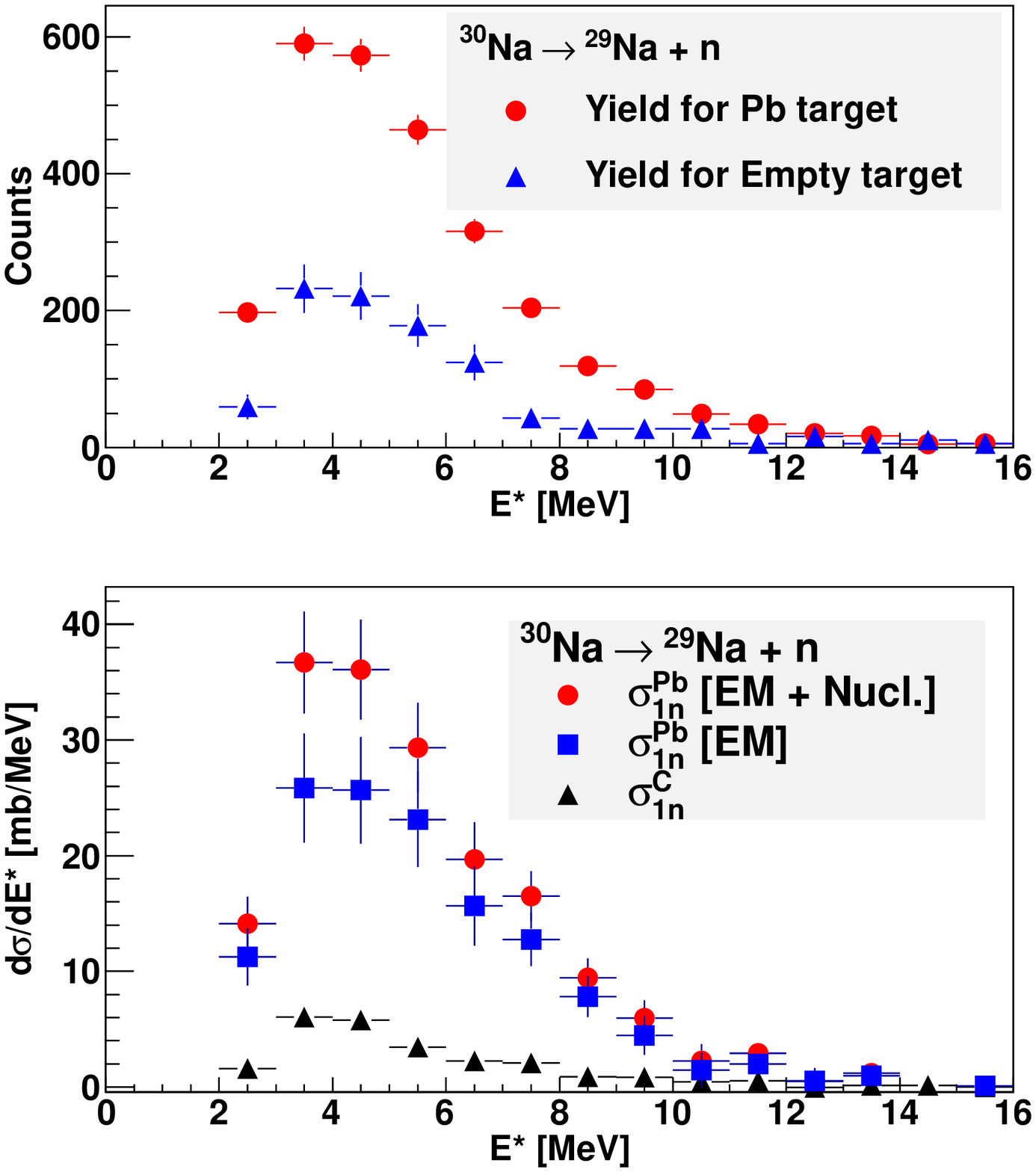}
\caption{(Top) The yields of  one neutron breakup reaction against excitation energy of $^{30}$Na for Pb target and without any target   
have been represented by filled circles and  triangles, respectively.
(bottom) One neutron breakup cross-sections against excitation energy of $^{30}$Na using  Pb-target (Coulomb + nuclear), C-target (nuclear) and
 pure Coulomb part have been denoted by circles, triangles and squares, respectively.}
\label{30na}
\end{figure}

The excitation energy $E^*$ of  $^{29,30}$Na were measured  using Pb and C targets.  The background contributions due to the reactions induced  by the
materials of the detectors  and air column were determined from the data taken without any  target and  this background data were subsequently 
subtracted from the data of Pb and C target.  Figure \ref{29na} (top)  shows reaction yields of one neutron breakup against the 
excitation energy of  $^{29}$Na  using Pb-target (filled circle) and without any target (filled triangle).
Similarly,  Figure \ref{30na} (top)  shows the reaction yields of one neutron breakup against $^{30}$Na  using Pb-target (filled circle) and 
without any target (filled triangle). The Coulomb dissociation (CD) cross section  of those neutron-rich nuclei using the $^{208}$Pb  target  (2.0 g/cm$^2$)  
was determined  after subtracting the nuclear contribution which was obtained  from the data with a $^{12}$C target (0.9 g/cm$^2$) 
with  proper scaling factor \cite{soft} (1.8).

With change of scaling factor of  10$\%$, the total cross-section of CD would change 1.4$\%$.
Filled circle, triangle and square in the lower panel of  Figure \ref{29na} and  Figure \ref{30na}, respectively show one neutron breakup reaction 
cross-section
of $^{29,30}$Na,  for Pb-target (Coulomb and nuclear),  C-target (nuclear) and  pure Coulomb part from Pb target,  respectively.
The CD cross section  for different core excited states can be further differentiated experimentally by the coincidence observation of
the characteristic $\gamma$-ray  of the core fragments with the  fragments and neutron \cite{ref e}. 
The situation of low-energy excitation in odd-z is very complex and the density of states at low-energy is high.
With the help of information available in the literature and simulation using those information along with present experimental  data,
the  detection efficiency  and feeding correction of the $\gamma$ rays  have been obtained. Utilising these efficiency factors, 
the CD cross-section for populating different core excited states can be deduced.
The sum of the partial CD cross-sections for populating the  excited states of  
$^{28}$Na upto 3.7MeV  is  30(4) mb. These cross-sections are  obtained from the Pb-target 
data  after subtracting its nuclear part. The cross-section for the ground state is  59(8) mb
 which is obtained from the difference between 
the total CD cross-section  and the cross-sections of the excited states.
Similarly, CD cross-section of $^{30}$Na  which populate various excited states of $^{29}$Na upto 4.7 MeV  is  46(6) mb.  
  The cross-section for the ground state is  121(14) mb which is obtained from the difference between 
the total CD cross-section  and the cross-sections of the excited states.

The measured invariant mass spectra  have been analyzed using the direct breakup model. The neutron-rich nucleus around N = 20,  
has been considered as a combination of the  core and a loosely bound valence neutron which may occupy one or more 
orbitals e.g. $s$, $d$, $p$ etc.. 
When the  projectile passes by a high Z target it  may be excited by absorbing the  virtual photons from the time dependent Coulomb 
field \cite{bar88}.                                                      
Hence due to this interaction the valence neutron moves from bound state to the continuum, while core behaves as a spectator.
Thus the nucleus breaks up into a neutron and the core. The electromagnetic breakup of loosely bound neutron-rich nuclei in energetic 
heavy ion collisions is dominated by dipole excitation due to smaller effective charge for higher multi-polarities \cite{bar01}.
Thus one neutron removal differential Coulomb dissociation cross section (CD)  for dipole excitations d$\sigma$/dE${^*}$ 
decomposes into an incoherent sum of components   d$\sigma (I^{\pi}_c$)/dE$^{*}$ corresponding to different core states 
$(I^{\pi}_c)$, populated after one neutron removal.
 For each core state, the cross-section further decomposes into incoherent sum over contribution
from different angular momenta j of the valence neutron in its initial state. Thus differential  cross-section  of  $^{29,30}$Na 
can be expressed through the following equation \cite{ref e}:

\begin{equation}
\frac{d\sigma (I^{\pi}_c)}{dE^*}=\ \frac{16\pi^3}{9\hbar c}N_{E1}(E^*)\displaystyle\sum_j C^2S(I^{\pi}_c,nlj)
 \\\times \displaystyle\sum_m|<q|(Ze/A)rY^l_m|\psi_{nlj}(r)>|^2 \
\end{equation}

Here, $N_{E1}(E^*)$ is the number of virtual photons as a function of excitation energy E* which can be computed
 adapting a semi-classical approximation \cite{bar88}. $\psi_{nlj}(r)$ and $<q|$ represent the single-particle wave 
function of the valence neutron in the projectile ground state (before breakup) and the final state wave-function of 
the valence neutron in the continuum (after breakup), respectively. The wave function of the outgoing neutron
 in the continuum is considered as a plane wave. $C^2S(I^{\pi}_c,nlj)$ represents the spectroscopic factor of 
the valence neutron  with respect to a particular core state $I^{\pi}_c$.

 Eq. (2) indicates that the dipole strength distribution is very sensitive to the single-particle wave
function which in turn depends on the orbital angular momentum and the binding energy of the valence neutron.
Thus, by comparing the experimental Coulomb dissociation cross section with the calculated one, information
on the ground state properties such as the orbital angular momentum of the valence nucleon and the corresponding
spectroscopic factor may be gained. The core state to which the neutron is coupled can be identified by
 the characteristic $\gamma$ -decay of the core after releasing the valence neutron. The
Coulomb breakup cross section which leaves the core in its ground state is obtained from the difference
between the total cross section and the excited-state(s) contribution(s).

In order to compare the experimental result with the  calculated data, one needs to convolute the instrumental
response function with the later. The instrumental response for a given value of E$^*$ has Gaussian distribution to a good
approximation \cite{Ref const, Ref noci,  haik}  at  excitation energy of 0.5 MeV or more  above neutron thresold 
but beyond that, it is asymmetric. 
The results of all the calculations,  shown in this article,  are
convoluted with the  response function. \\

\section{Result}

\begin{figure*}[t]
\centering
\includegraphics[width=16 cm]{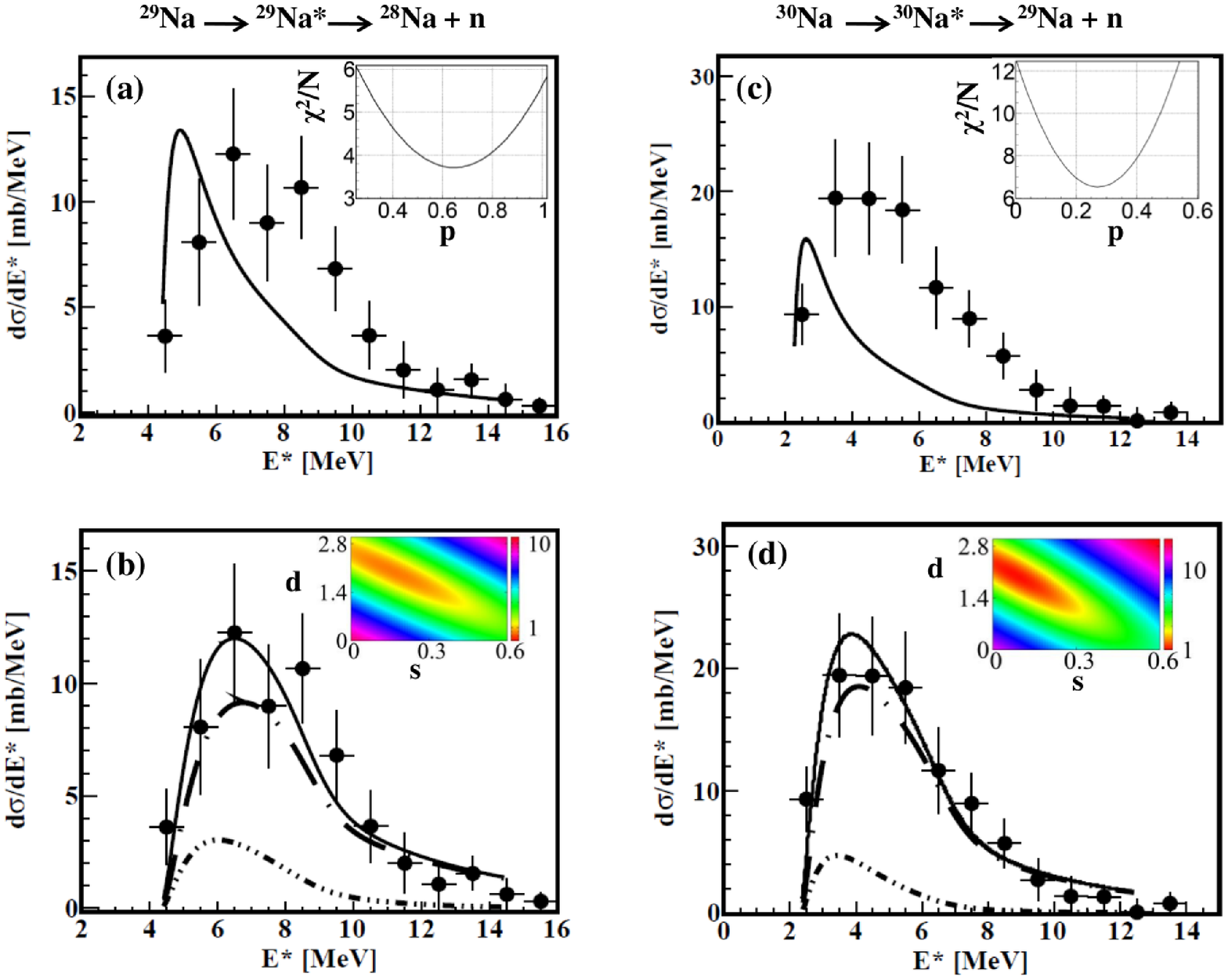}%
\caption{\label{all_fit}
(a-b)Experimental differential pure Coulomb dissociation cross-section of $^{29}$Na, breakup into $^{28}$Na (gr.) and  one neutron. 
 The solid line represents differential CD cross-section using  direct breakup model where valence neutron is occupying p-orbital, 
or combination of s and d-orbital, respectively. The dashed  and dotted-dashed line represent the calculated CD cross-section with
valence neutron in $d$ and $s$ orbital  with respective spectroscopic factors. (c-d) Experimental differential 
pure Coulomb dissociation cross-section  of $^{30}$Na  
against excitation energy and the solid line represents    differential CD cross-section using  direct breakup model where the 
 valence neutron is occupying  $p$-orbital, or combination of $s$ and $d$-orbital, respectively. The dashed  and dotted-dashed line 
represent the calculated CD cross-section with  $d$ and $s$ wave components. The inset of every figures  show 
the $\chi^2/N$ of the fitting between experiment and calculated one  against the spectroscopic factor for the
 valence neutron of that orbital.}
\end{figure*}

 The total Coulomb dissociation cross section for  $^{29}$Na into  $^{28}$Na and one neutron amounts to 
89 $( 7)$ mb, after  integration up to 10 MeV excitation energy. No resonance-like structure has been observed.
The data analysis for $^{29}$Na  shows that the major  part (∼ 67$( 11)$ $\%$) of the breakup cross section leaves 
the core  $^{28}$Na in its ground state and approximately  ($\sim$33( 5)$\%$) of  the  fragments are found in the excited
 states which  could be deduced from the invariant mass spectra,  obtained through coincidence of the  $\gamma$-ray sum spectra with the 
 core fragment (i.e., $^{28}$Na) and one neutron.  Left panel of the Figure \ref{all_fit} shows the experimental  
 differential Coulomb dissociation cross section with  respect to the  excitation energy (E$^*$) of $^{29}$Na 
 which breaks up into a neutron and a $^{28}$Na fragment in  its ground state  (filled circles). 
The spectrum was obtained after subtracting  contribution due to the excited states from the total 
differential cross-section of pure Coulomb breakup of $^{29}$Na. 
The valence neutron of $^{29}$Na  is loosely bound (S$_n$ = $4.4$ MeV). Considering the nucleus as a core 
and loosely bound neutron, the calculation for CD cross-section using the direct breakup model  \cite{ref e} 
have been performed using various valence neutron occupying orbitals. The outgoing neutron in the continuum is approximated by  a plane-wave.
 To understand the valence  neutron occupation orbital with probability, the experimental 
Coulomb Dissociation cross-section d$\sigma (I^{\pi}_c$)/dE$^{*}$ of $^{29}$Na  into  ground state of $^{28}$Na and 
one neutron  have been  compared with direct breakup model calculation.  All experimental  d$\sigma$/dE$^{*}$ distributions are 
shown in figures without  acceptance and efficiency corrections
 for the neutron detector. Instead, the calculated cross  sections were convoluted with the detector response 
obtained from detailed simulations.  Comparison between the experimental data on Coulomb dissociation with
 direct breakup model calculation  using  $p$,  and combination of $s$ and $d$ orbital, respectively are shown 
in the figures \ref{all_fit}a, \ref{all_fit}b, respectively. The experimentally observed shape  of  the spectrum 
is in good agreement with the calculated one considering valence neutron  in $s$ and $d$ orbital. 
The solid curve in the figures \ref{all_fit}a, \ref{all_fit}b represent the calculated d$\sigma (I^{\pi}_c$)/dE$^{*}$
 using the direct-breakup model with the valence neutron in the $p$ and  combination of $s$ and $d$ orbitals, respectively. 
The inset of the  Figures \ref{all_fit}a show  the $\chi^2/N$ for fitting against the spectroscopic factors of 
the valence neutron in $p$ orbital. The inset of the  Figure \ref{all_fit}b, shows   three dimensional plot  of  the $\chi^2/N$  
for fitting (z-axis in color shade) against the spectroscopic factors of the valence neutron in the $s$, $d$ orbital (x and y axis).  
 It is evident from  Figure \ref{all_fit}b that the best fit of the experimental data can be 
obtained with  calculation where the valence neutron is occupying  
 combination of the $s$ and $d$ orbitals. The $\chi^2/N$ for  the fit suggests that the neutron is 
 occupying the  $s$ and $d$ orbitals with the spectroscopic factors $0.07(7)$ and $2.1(3)$, 
 respectively. The errors quoted in the spectroscopic factors are obtained from the $\chi^2$  distribution and 
the errors  are one sigma  i.e, within $68\%$  confidence limit.  
The dashed  and dotted-dashed line in Figure  \ref{all_fit}b represent the calculated CD cross-section with 
valence neutron in the $d$ and $s$ orbital with above mentioned spectroscopic factors. The shaded region in the figure reprsents 
the error associated with fitted  curve of calculation. This   error is  associated with spectroscopic factor, 
obtained from fitting with experimental data.

The total Coulomb dissociation cross section for  $^{30}$Na into  $^{29}$Na and a neutron amounts to 167 $(13)$ mb,
after  integration  up  to 10 MeV excitation energy. No resonance-like structure has been observed.
 The data analysis for  $^{30}$Na shows that the major  part (∼ 72 $( 10)\%$) of the breakup cross section 
leaves the core  $^{29}$Na in its ground state and approximately  ($\sim$28 (4)$\%$) of  the  fragments are 
found in the excited states. Right panel of Fig. \ref{all_fit} shows the experimental  
 differential Coulomb dissociation cross section with 
respect to the  excitation energy (E$^*$) of $^{30}$Na  breaking up into a neutron and a $^{29}$Na fragment in 
its ground state  (filled circles). This spectra was obtained after subtracting excited state contribution 
from the spectra of total differential cross-section of pure Coulomb breakup of $^{30}$Na. 
The valence neutron of $^{30}$Na  is sufficiently loosely bound (S$_n$ = $2.37$ MeV).  The experimental 
Coulomb Dissociation cross-section (d$\sigma$/dE$^{*}$) of $^{30}$Na into 
 ground state of $^{29}$Na and one neutron  have been  compared with direct breakup model 
calculation   considering the  valence neutron in the $p$,  or combination of  $s$ and $d$ orbitals.
 Inset of the figure shows $\chi^2$/N obtained  from fitting  with variation of spectroscopic factors.
The data can be well reproduced by a fit including contributions from the wave functions involving $l$ = $0$ and $l$ = $2$
neutrons, as shown in Fig. \ref{all_fit}d. The inset of the Fig. \ref{all_fit}d shows
spectroscopic factors  of  $s$ and $d$ orbital occupation of the valence neutron corresponding to
distribution  of the $\chi^2/N$.  The spectroscopic factors obtained from the fit to the
data using the plane-wave approximation for the  neutrons occupying the $s$ and $d$ orbitals
are $0.05(5)$ and $2.03(30)$,  respectively.  The  dashed  and dotted-dashed line in Figure  \ref{all_fit}d 
represent the calculated CD cross-section using the  valence neutron in $d$ and $s$ orbital with 
 respective spectroscopic factors.The ground state spin and parity of $^{30}$Na from this experimental results favor
3/2$^+ \otimes$3/2$^+$ i.e. either 3$^+$ or 2$^+$ or  1$^+$. The shaded region in the figure reprsents 
the errors associated with fitted  curve of calculation. This   error is  associated with spectroscopic factor, 
obtained from fitting with experimental data.

\begin{table*} 
\centering
 \caption{Coulomb dissociation cross sections of $^{29,30}$Na for  various core state and valence-neutron orbitals. 
Cross sections  obtained from the direct-breakup model for neutrons occupying $s$ an $d$ orbitals with a spectroscopic
 factor of one are given for comparison. The cross sections are integrated up to 10 MeV for $^{29}$Na and $^{30}$Na. 
The corresponding spectroscopic factors from shell-model calculations and the ones derived from the experiment 
are quoted in the last two columns.}
\vspace{5mm}
 \label{tab:1}

\begin{tabular}{|ccccccc|}
\hline
Isotope & Core  & Neutron  &Cross section (mb) &Spectroscopic factor&&\\
&&&&&&\\
        &state &   orbital  & Expt. & Expt&&\\ 
        & $I^{\pi}_c(E_c; $[MeV]) &           &     &&&\\
&&&&&&\\
\hline
 $^{29}Na$&  $1^+$ (0.0)& &$59(8)$&&&\\
& & $1s$ &  &$0.07(7)$& &\\
&  & $0d$&&$2.1(3)$ &&\\
&$0<E_c<3.7$&&30($4$)&&&\\
\hline
&&&&&&\\
$^{30}Na$&$3/2^+$ (0.0) &&$121(14)$ &&& \\
& & $1s$ && $0.05(5)$&&\\
& &  $0d$&&$2.03(30)$&&\\
&$0<E_c<4.7$&&46$(6)$&&&\\
\hline 
\end{tabular}
\end{table*}

\section{Discussion}
\label{}

The ground state spin and parity of $^{29,30}$Na were measured by magnetic resonance \cite{huber}. But no detailed 
measurement on  ground state configuration is available.  $\gamma$-ray spectroscopy data \cite{Ref hurst, Ref tripathi1,Ref tripathi2,isol}
 were interpreted  as  intruder dominated ground state configuration.  Present experimental 
CD cross sections of $^{29,30}$Na along with the  calculated  one from the direct breakup model
have been summarized in the table \ref{tab:1}. The  Coulomb breakup calculation  are compared with
the experimental findings.  The dominant ground state configuration of  $^{29}$Na is $^{28}$Na$_{gs}(1^+)\otimes\nu_{s,d}$.
 The ground state spin and parity of $^{29}$Na from this experiment  favor 1$^+ \otimes$3/2$^+$ i.e, either 5/2$^+$, 3/2$^+$ or  1/2$^+$. 
A comparison between the spectroscopic factors obtained from this experimental data with shell model calculations  using 
various interaction have been presented in the   table \ref{tab:1}. In USD-B calculation the valence space is composed of the 
$sd$ shell for both protons and  neutrons.  For $sdpf$-M,  the valence space is composed of the $sd$ shell for protons  and   
allows mixing between   $sd$ and $pf$ shell orbitals for neutrons. The Hamiltonian has been recently introduced \cite{er shell}. 
The calculated spectroscopic factors are given in table \ref{tab:1}.  Both USD-B \cite{brown} shell model calculation, and  
MCSM shell model calculations \cite{Ref utsuno} favor 3/2$^+$ as the ground state spin and parity of $^{29}$Na. This is in agreement 
with the earlier experimentally measured value \cite{huber}. 
Utsuno et al. \cite{Ref utsuno} showed that  the measured two neutron separation energy, magnetic and electric moments of
$^{29}$Na can be  explained by both USD and $sdpf$-M. The excited core contributions in their ground state configuration are 
around $\sim33(5)\%$ as per our experimental  observation for  $^{29}$Na isotopes and  the core  excited state 
contribution above 2.0 MeV is around 27$\%$.  To a simple approximation, if this amount is considered  wholly due to  $2p-2h$ 
configurations, then according to MCSM \cite{Ref utsuno} calculations, one can obtain $\sim$ 4.0 MeV as lower limit of the $sd-pf$ shell 
gap in $^{29}$Na. It is clear from Figure \ref{all_fit}b that the experimental spectroscopic factor for the valence neutron in the $d$ 
orbital coupled with $^{28}$Na$_{gs}(1^+)$ is  2.1 $(3)$  and this is in good agreement with
USD-B shell model calculation (2.18).  Earlier it was shown that the measured mass of $^{29}$Na could be explained
 by sd-shell model calculation \cite{wild}.

 The ground state spin and parity of $^{30}$Na has been measured earlier as being $2^+$ \cite{huber} and this value can be reproduced by both 
 USD \cite{brown} shell model and MCSM shell model calculations \cite{Ref utsuno}.  No experimental data on ground state configuration of 
this nucleus is available.  To explain measured reduced matrix element of this isotope, it has been considered that the ground state is pure
two-particle-two-hole deformed ground state \cite{isol}. But  present experimental data favors  $^{29}Na_{gs}(3/2^+)\otimes\nu_{s,d}$ as 
major ground state configuration  of $^{30}$Na ($N$ = $19$)  and  the core $^{29}$Na excited state contribution  i.e, particle hole 
configurations in the ground state is around 46 $(6)$ mb ($28(4)\%$ of total CD cross-section). If it is  approximated that the excited 
states contributions are  entirely due to  $2p-2h$ configurations then a comparison of our experimental findings with shell-model 
calculation  using the MCSM  \cite{Ref utsuno} suggests a  lower limit of the $sd-pf$ shell gap, around $4.3$ MeV in this nucleus.  
For the first time experimental  quantitative spectroscopic  factors of the valence neutron in the $s$ and $d$ orbital   have been measured.
Unlike, $^{29}$Na, for $^{30}$Na, the experimental spectroscopic  factor for occupation of the $d$-orbital (2.03)
 deviates from $sd$-shell (USD-B) calculation (2.97).  Hence reduced spectroscopic factor of valence neutron occupying 
$d$-orbital could be due to particle hole configuration across the reduced the $sd-pf$ shell gap. 
Since ground state spin and parity of this neutron-rich nucleus is $2^+$, the valence neutron occupying $pf$ orbital, should be coupled with 
negative parity excited states of $^{29}$Na. So far very little is  known about the excited states of $^{29}$Na. On the other hand,
 Coulomb breakup cross-section using direct breakup model for valence neutron in $f_{7/2}$ orbital coupled with 
core excited state of $1.5$ MeV is around $21$ mb and  it further reduced to 10 mb when coupled with 3.0 MeV 
excited core state. When valence neutron is occupying $p$ orbital, the direct breakup model calculation for same situation is
108 mb and 67 mb, respectively. Considering experimental excited state CD cross-section, 46 $( 6 )$mb, it may be possible
 that valence neutron across the shell gap is occupying either pure $f$  or $p$ orbital or mixing of the $p$ and $f$ orbital.
 
In a nut-shell,  present experimental data of Coulomb breakup suggests that ground state properties of $^{29}$Na can be 
explained by USD-B shell model calculation. But the situation is different for that of $^{30}$Na. The spectroscopic factor for 
valence neutron in the $d$ orbital is  almost 1/3 reduced compared to USD-B calculation. This could be tentatively, for particle
 hole configuration  across the shell gap and valence neutron may occupy either   $f$ or $p$ orbital or mixing of both the $pf$ 
orbital. Thus, it may be concluded that boundary of so-called island of inversion has been started from $^{30}$Na, instead of $^{29}$Na. 
Wildenthal et al. \cite{wild}, showed that measured mass of $^{29}$Na  can be explained by USD shell model calculation  
  but the same for  $^{30}$Na  was not fully reproduced by sd-shell model calculation.
The  measured B(E2; 2$^+ \rightarrow 3^+)$  value  for  $^{30}$Na \cite{Ref c, Ref cp} is  around 30$\%$ lower
than the  value calculated by MCSM with the $sdpf$-M interaction.  However more details theoretical calculation is necessary to 
understand nucleon-nucleon interaction for this neutron-rich nuclei near the so-called island of inversion'. 
  
\section{Conclusion}
First results on Coulomb breakup measurements of the neutron-rich  $^{29,30}$Na  nuclei, at
energies of 400-430 MeV/nucleon has been reported. The observed low-lying dipole strength in these
 neutron-rich Na isotopes can be understood as a direct-breakup mechanism and no resonance like structure has been observed.
The shape of the experimental  differential Coulomb dissociation cross section and its comparison with the calculated
cross section  suggests  the predominant ground-state configuration  as $^{28}$Na${(1^+)}\otimes\nu_{s,d}$ and
$^{29}$Na${(3/2^+)}\otimes\nu_{ s,d}$ for $^{29}$Na and $^{30}$Na  respectively. The  ground state spin and 
parity of these nuclei, obtained from present measurement  are in agreement with earlier reported values.
Thus the first results on ground state configurations and qualitative spectroscopic information of valence neutron
 occupying the $s$, $d$ orbitals, obtained via direct method have been reported in this letter. According to this experimental
results, the valence neutron is occupying mainly $d$ orbital for both the neutron-rich Na nuclei ($N$ = $18,19$).
But experimentally obtained  spectroscopic factor 2.1 $(3$) for valence neutron in the $d$ orbital  of $^{29}$Na  
is in closer agreement with modified  $sd$-shell (USD-B) calculation ($2.18$). On the other hand the same for $^{30}$Na  
is different and experimentally obtained spectroscopic factor for valence neutron in the $d$ orbital is much reduced 
2.03 $(30)$  compared to  sd-shell (USD-B) calculation ($2.97$). This could be due to particle-hole excitation
of the valence neutron across the shell-gap.  A comparison of our experimental findings on the core 
excited states contributions in the ground state configuration with the shell-model calculation using the MCSM suggests a 
lower limit of the $sd-pf$ shell gap in $^{30}$Na of around $4.3$ MeV. Thus present experimental data  pointed that   $^{30}$Na 
 is the boundary of ``island  of inversion'', instead of $^{29}$Na. However more detail theoretical calculation is necessary to 
understand the nucleon-nucleon interaction for this neutron-rich nuclei near so-called island  of inversion.

\section*{Acknowledgement}
We are thankful to the accelerator people of GSI for their active support during the experiment. Authors are thankful to
 Prof. B.A. Brown, Michigan state University for providing us shell model calculation and 
Prof. Sudeb  Bhattacharya, Kolkata for critically reviewing the manuscript and suggestions. Author, Ushasi  Datta  
acknowledges   Alexander von Humboldt foundation, Germany  and SEND project  grants (PIN:11-R$\&$D-SIN-5.11-0400) from 
 Department of Atomic Energy (DAE), Govt. of India  for   financial support  for work.

\end{document}